\documentclass[reqno]{article}
\newcommand{\ds}{\displaystyle}
\newcommand{\dsf}{\ds\frac}

\newcommand{\beq}{\begin{equation}}
\newcommand{\eeq}{\end{equation}}

\usepackage{ae} 
\usepackage[T1]{fontenc}
\usepackage[ansinew]{inputenc}
\usepackage{amsmath}
\usepackage{graphicx}
\usepackage{color}
\usepackage[colorlinks]{hyperref}
\usepackage{multicol}
\begin{document}
\footnotesize

\begin{center}
\bf Magnetic instability due to flux avalanches in superconductors
\end{center}

\begin{center}
N. A. Taylanov and M. Samadov
\end{center}

\begin{center}
\emph{National University of Uzbekistan, Uzbekistan}
\end{center}
\begin{center}
(\emph{Dated: December 16, 2014})
\end{center}

\begin{center}
 \textcolor{black}{Abstract}
\end{center}
We have studied theoretically the space-time evolution of the
thermal and electromagnetic perturbation in a superconductor with
the linear current-voltage characteristic in the flux flow regime.
On the basis of a linear analysis of a set of differential
equations describing small perturbations of temperature and
magnetic field we will found that under some conditions an
instability may occur in the sample, which takes into account an
inertial properties of the vortices mass.

\vskip 0.5cm


\begin{multicols}{2}{Bean's critical state with its spatially nonuniform
flux distribution is not at equilibrium and under certain
conditions the smooth flux penetration process becomes unstable
[1-5]. The spatial and temporal development of this instability
depends on the sample geometry, temperature, external magnetic
field, its rate of change and orientation, initial and boundary
conditions, etc. Instabilities in the critical state result in
flux redistribution towards the equilibrium state and are
accompanied by a significant heat release, which often leads to
the superconductor-to-normal-transition. The basic instability
observed in Bean's critical state is the flux jump instability,
which was discovered already in the early experiments on
superconductors with strong pinning [4].

Let us assume that a small perturbation of temperature or flux
occurring in Bean's critical state. This perturbation can be caused
by an external reason or a spontaneous fluctuation arising in the
system itself. The initial perturbation redistributes the magnetic
flux inside the sample. This flux motion by itself induces an
electric field which leads to dissipation. This additional
dissipation results in an extra heating which in turn leads to an
additional flux motion. This "loop" establishes a positive feedback
driving the system towards the equilibrium state. The flux jump
instability exhibits itself as suddenly appearing flux avalanche and
Joule heat release [4, 5].

Recently, Chabanenko et all. [6] have reported an interesting
phenomenon in their experiments  -  convergent oscillations of the
magnetic flux arising from flux jump avalanches. The authors
argued that the observed oscillations due to flux avalanches can
be interpreted as a result of the existence of a definite value of
the effective vortex mass. Thus, it is necessary to take into
account collective modes, i.e., the inertial properties of the
vortices in studying the dynamics of the flux avalanches.  Prior
to the jump, the mixed state of superconductors is characterized
by nonuniformly distributed magnetic induction localized near the
surface. As a result of the avalanche, the flux rushes from either
sides of the sample towards the center [6]. Two fronts of the
penetrating flux collide in the center of the sample and, owing to
the existing vortex mass, give rise to the local surplus density
of the magnetic flux that exceeds the value of the external
magnetic field. The repulsion force in the vortex structure at the
center of the sample that have resulted from its compaction,
initiates the wave of the vortex density of the inverse direction
of propagation. Upon reaching the surface, this wave is reflected
from it. This results in the oscillations in the vortex system
[6]. The limitation of the number of oscillations observed is
caused by the existence of damping. One succeeds in observing the
oscillation of the vortex density only owing to a strong
compression of the vortex structure as a result of the giant
avalanche-flux [6-11].

One important parameter characterizing a quantized vortices is the
effective (intrinsic) mass, which can be associated with its
motion [12-21]. Due to its importance, the concept of the vortex
mass was discussed extensively over the years but remains a
controversial issue. One point of view is that mass plays no role
in the dynamics since an inertial term in the equation of motion
of a vortex is always negligible next to the viscous drag force.
Experimentally, the presence of an inertial term is difficult to
detect since at low temperatures vortices in superconductors are
pinned, and if they move at all, their motion is dominated by
viscosity. We are aware of only one such attempt, with
inconclusive results8. It is quite clear that in order to check if
an inertial term plays a significant role in vortex dynamics, one
should realize conditions where the viscous drag force is not
dominant. Our simulations will be performed in a region near the
center of the sample where the currents vanish and this force is
small [15].

In the present work, we study the the dynamics of the magnetic
flux avalanches which takes account inertial properties of the
vortex matter. It is shown that at under some conditions vortex
instability can be observed during the flux avalanche process.

Bean [1] has proposed the critical state model which is
successfully used to describe magnetic properties of type II
superconductors. According to this model, the distribution of the
magnetic flux density $\vec B$ and the transport current density
$\vec j$ inside a superconductor is given by a solution of the
equation

\begin{equation}
rot\vec B=\dsf{4\pi}{c}\vec j.
\end{equation}
When the penetrated magnetic flux changes with time, an electric
field $\vec E(r, t)$ is generated inside the sample according to
Faraday's law

\begin{equation}
rot\vec E=\dsf{1}{c}\dsf{d\vec B}{dt}.
\end{equation}
In the flux flow regime the electric field $\vec E(r, t)$ induced
by the moving vortices is related with the local current density
$\vec j(r, t)$ by the nonlinear Ohm's law

\begin{equation}
\vec E=\vec v\vec B.
\end{equation}
To obtain quantitative estimates, we use a classical equation of
motion of a vortex, which it can derived by integrating over the
microscopic degrees of freedom, leaving only macroscopic forces
[21]. Thus, the equation of the vortex motion under the action of
the Lorentz, pinning, and viscosity forces can be presented as

\begin{equation}
m\dsf{dV}{dt}+\eta V+F_L+F_p=0.
\end{equation}
Here $\mu$ is the vortex mass per unit length, $\vec
F_L=\dsf{1}{c}\vec j\vec \Phi_0$ is the Lorentz force,  $\vec
F_p=\dsf{1}{c}\vec j_c\vec \Phi_0$,
$\eta=\dsf{\Phi_0H_{c2}}{c^2\rho_n}$ is the flux flow viscosity
coefficient, $\Phi_0=\pi hc/2e$ is the magnetic flux quantum,
$H_{c2}$ is the upper critical field of superconductor, $\rho_n$
id the normal state resistivity, $j_c$ is the critical current
density [4]. For simplicity we have neglected the Magnus force,
assuming that it is much smaller then the viscous force (for
example, for Nb see, [6]). In the absence of external currents and
fields, the Lorentz force results from currents associated with
vortices trapped in the sample.

In combining the relation (3) with Maxwell’s equation (2), we
obtain a nonlinear diffusion equation for the magnetic flux
induction $\vec B(r, t)$ in the following form
\begin{equation}
m\dsf{dV}{dt}+\eta V=-\dsf{1}{c}\Phi_0(j-j_c),
\end{equation}
\begin{equation}
\dsf{d\vec B}{dt}=\nabla [\vec v\vec B].
\end{equation}

The temperature distribution in superconductor is governed by the
heat conduction diffusion equation

\begin{equation}
\nu (T)\dsf{dT}{dt}=\nabla[\kappa(T)\nabla T]+\vec j\vec E,
\end{equation}
Here $\nu=\nu(T)$ and $\kappa=\kappa(T)$ are the specific heat and
thermal conductivity, respectively. The above equations should be
supplemented by a current-voltage characteristics of
superconductors, which has the form

$$
\vec j=j_{c}(T, \vec B, \vec E).
$$
In order to obtain analytical results of a set Eqs. (5)-(7), we
suggest that $j_c$ is independent on magnetic field induction $B$
and use the Bean critical state model $j_c=j_c(B_e, T)$, i.e.,
$j_c(T)=j_0-a(T-T_0)$ [1];  where $B_e$ is the external applied
magnetic field induction, $a=j_0/(T_c-T_0)$, $T_0$ and $T_c$ are
the equilibrium and critical temperatures of the sample,
respectively, $j_0$ is the equilibrium current density. For the
sake of simplifying of the calculations, we perform our
calculations on the assumption of negligibly small heating and
assume that the temperature profile is a constant within the
across sample and thermal conductivity $\kappa$ and heat capacity
$\nu$ are independent on the temperature profile [5].

We study the evolution of the thermal and electromagnetic
penetration process in a simple geometry - superconducting
semi-infinitive sample $x\geq 0$. We assume that the external
magnetic field induction $B_e$ is parallel to the z-axis and the
magnetic field sweep rate $\dot{B_e}$ is constant. When the
magnetic field with the flux density $B_e$ is applied in the
direction of the z-axis, the transport current $\delta j(x, t)$
and the electric field $\delta E(x, t)$ are induced inside the
slab along the y-axis. For this geometry the spatial and temporal
evolution of thermal and magnetic field perturbations
\begin{equation}
\begin{array}{l}
T=T_0+\Theta(x, t),\\
\quad\\
B=B_e+b(x, t),\\
\quad\\
V=V_0+v(x, t)\\
\end{array}
\end{equation}
where $T_0(x)$, $B_e(x)$ and $V_0(x)$ are solutions to the
unperturbed equations, which can be obtained within a
quasi-stationary approximation; are described by the following
system of differential equations [8, 11]

\begin{equation}
\dsf{d\Theta}{dt}=2v-\beta\Theta,
\end{equation}
\begin{equation}
\mu\dsf{dv}{dt}+v=-\dsf{db}{dx}+\beta\Theta,
\end{equation}
\begin{equation}
\dsf{db}{dt}=\left(\dsf{db}{dx}+b\right)+\left(\dsf{dv}{dx}+v\right),
\end{equation}
where we have introduced the dimensionless parameters
$$
\mu=\dsf{c\Phi_0}{4\pi\eta^2}\dsf{B_e}{2L^2}, \quad
\beta=\dsf{4\pi}{c}\dsf{j_{c}^{2}L^2}{\nu(T_c-T_0)}.
$$
$$
b=\dsf{B}{B_e}=\dsf{c}{4\pi}\dsf{B}{j_cL},\quad
\Theta=\dsf{4\pi}{c}\dsf{2\nu}{B_{e}^{2}}, \quad v=V\dsf{t_0}{L},
\quad L=\dsf{c}{4\pi}\dsf{B_e}{j_{c}}.
$$
variables

$$
z=\dsf{x}{L},\quad
\tau=\dsf{t}{t_0}=\dsf{c\Phi_0}{4\pi\eta}\dsf{B_e}{2\mu_0j_cL^2}t,
$$

Here $L$ is the magnetic field penetration depth, which is
determined from equation (3)
\begin{equation}
B(x, t)=B_e+\dsf{4\pi}{c}j_c(x-L),
\end{equation}
with the appropriate boundary conditions
\begin{equation}
dB(0, t)=B_e,\quad B(L, t)=0.
\end{equation}

Assuming that the small thermal and magnetic perturbations has
form $\Theta(x,t), b(x,t),  v(x,t)\sim\exp[\gamma t],$ where
$\gamma$ is the eigenvalue of the problem to be determined, we
obtained from the system Eqs. (9)-(11) the following dispersion
relations to determine the eigenvalue problem

\begin{equation}
(\gamma+\beta)
\dsf{d^2b}{dx^2}-[(\gamma+\beta)\mu-2\beta]\dsf{db}{dx}+[(\mu+1)\gamma^2+[(\mu-1)\beta-
\mu-1]\gamma-(\mu-1)\beta]b=0
\end{equation}
The instability of the flux front is defined by the positive value
of the rate increase Re $\gamma$>0. The instability occurs at the
condition

The instability of the flux front is defined by the positive value
of the rate increase Re $\gamma$>0.

An analysis of the dispersion relation shows that, the grows rate
is positive Re $\gamma$>0, if $\mu>\mu_c=2$ and any small
perturbations will grow with time. For the case when $\mu<\mu_c$,
the growth rate is negative and the small perturbations will
decay. At the critical value of $\mu=\mu_c$, the growth rate is
zero $\gamma$=0. For the specific case, where $\mu=1$ the growth
rate is determined by a stability parameter $\beta$. Thus, the
stability criterion can be written as

$$
\beta>1.
$$
For the case, where thermal effects is negligible ($\beta=1$) we
may obtain the following dispersion relation
\begin{equation}
\dsf{d^2b}{dx^2}-\mu\dsf{db}{dx}+(\gamma-1)(\mu+1)b=0.
\end{equation}
Seeking for $b\sim\exp(ikx)$ in dispersion relation, the growth
rate $\gamma$ dependence can be obtained as a functions of
wavenumber k.

\begin{center}
\includegraphics[width=3in]{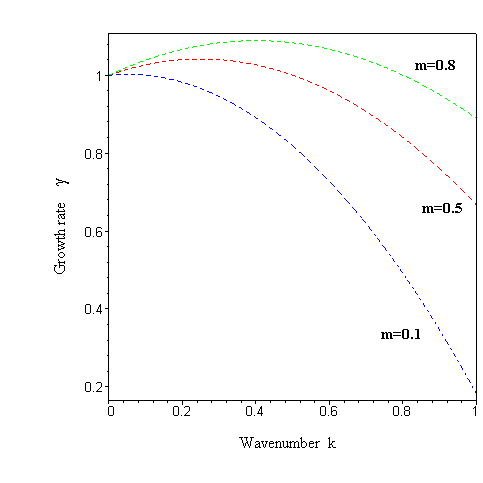}
\end{center}
\begin{center}
Fig.1.  The dependence of the growth rate on the wavenumber for
$\mu=0.1, 0.5, 0.6$.
\end{center}

The stability of the system depends on the growth rate, $\gamma$,
given in (15). We analyze the growth rate of small perturbations
as a function of wavenumber k. When $k<k_c=\mu$ the growth rate is
positive and any small perturbations will grow with time. For wave
number $k>k_c$, the growth rate $\gamma$ is negative.
Consequently, the small perturbations always decay. It can be
shown that, for wave number $k=k_c$ the growth rate is zero
$\gamma=0$. As the wave number approaches zero $k\longrightarrow
0$ or infinity $k\longrightarrow \infty$ the growth rate
approaches $\gamma=1$ and small perturbations grow with time. As
the wave number approaches unity $k=1$ the growth rate is
determined by the value of $\mu$
$$
\gamma=\dsf{2\mu}{\mu+1}.
$$
For $\mu=0$ the growth rate is zero $\gamma=0$. For $\mu=1$ the
growth rate is unity $\gamma=1$. Since the growth rate is zero at
the critical wave number and approaches to unity in the limit of
zero wave number, there must exist a wave number in between that
maximizes the growth rate. Fig. (1-4) shows the growing rate,
$\gamma$, as a function of the wave number k, for various values
mass $\mu$. As the value of $\mu$ increases, the corresponding
growth rate increases.

In the present work, we study the spatial and temporal evolution
of small thermal and magnetic perturbation in type-II
superconductor sample in the flux flux regime, assuming that an
applied field parallel to the surface of the sample. On the basis
of a linear analysis of a set of differential equations describing
small perturbations of temperature and magnetic field we will
found that under some conditions an instability may occur in the
sample, which takes into account an inertial properties of the
vortices mass.

\vskip 0.5cm

\begin{center}
{\bf  References}
\end{center}

\vskip 0.5cm
\begin{enumerate}
\item C. P. Bean, Phys. Rev. Lett., 8, 250, 1962; Rev. Mod. Phys.,
36, 31, 1964.

\item P. S. Swartz and S. P. Bean, J. Appl. Phys., 39, 4991, 1968.

\item S. L. Wipf, Cryogenics, 31, 936, 1961.

\item  R. G. Mints, and A. L. Rakhmanov, Rev. Mod. Phys., 53, 551,
1981.

\item R. G. Mints and A. L. Rakhmanov, Instabilities in
superconductors, Moscow, Nauka, 362, 1984.

\item V. V. Chabanenko, V. F. Rusakov, V. A. Yampol’skii, S.
Piechota, A. Nabialek, S. V. Vasiliev, and H. Szymczak, arXiv:
cond-mat/0106379v2, 2002.

\item S. Vasiliev, A. Nabialek, V. Chabanenko, V. Rusakov, S.
Piechota, H. Szymczak, Acta Phys. Pol. A 109, 661, 2006.

\item A. Nabialek, S. Vasiliev, V. Chabanenko, V. Rusakov, S.
Piechota, H. Szymczak, Acta Phys. Pol. A, 114, 2008.

\item S. Vasiliev, A. Nabialek, V. F. Rusakov, L. V. Belevtsov,
V.V. Chabanenko and H. Szymczak,  Acta Phys. Pol. A, 118, 2010.

\item V. Rusakov, S. Vasilieva, V.V. Chabanenko , A. Yurov, A.
Nabialek, S. Piechotaa and H. Szymczak, Acta Phys. Pol. A, 109,
2006.

\item V. V. Chabanenko, V.F. Rusakov , A. Nabialek , S. Piechota ,
S. Vasiliev , H. Szymczak, Physica C, 369, 2002.

\item N. H. Zebouni, A. Venkataram, G. N. Rao, C. G. Grenier, J.
M. Reynolds, Phys.Rev.Lett., 13, 606, 1964.

\item H. Suhl, Phys.Rev.Lett., 14, 226, 1965.

\item H. T. Coffey, Cryogenics, 7, 73, 1967.

\item N. V. Kopnin. Pis’ma v ZhETF 27, 417, 1978.

\item G. Baym, E. Chandler. J. Low Temp. Phys., 50, 57, 1983.

\item E. B. Sonin, V. B. Geshkenbein, A. van Otterlo, G. Blatter.
Phys. Rev. B 57, 575, 1998.

\item M. J. Stephen, J. Bardin. Phys. Rev. Lett., 14 112, 1965.

\item G. E. Volovik, Pis’ma v ZhETF, 65, 201, 1997.

\item E. M. W. Coffey, Phys. Rev. B, 49, 9774, 1994.

\item J. I. Gittleman, B. Rosenblum. Journ. of Appl. Phys., 39,
2617, 1968.

\newpage

\end{enumerate}
}\end{multicols}
\begin{center}
\includegraphics[width=3in]{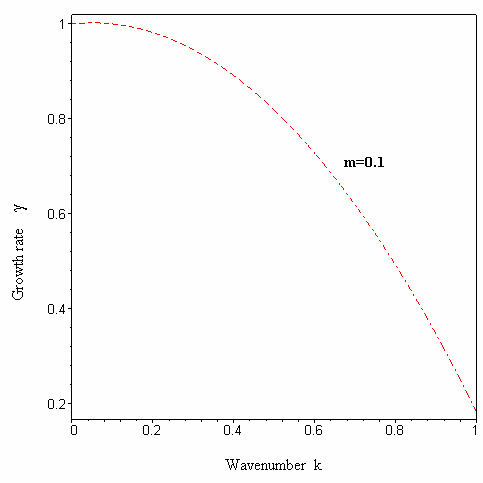}
\includegraphics[width=3in]{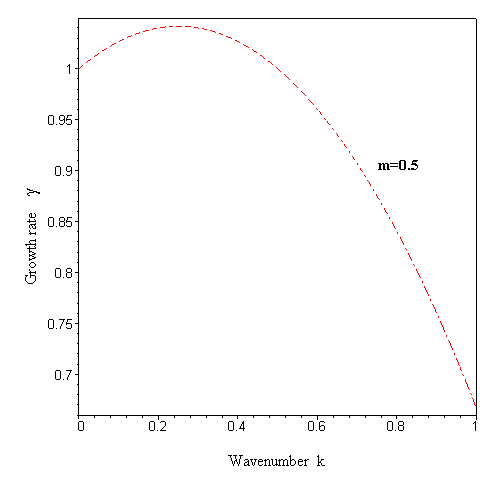}
\includegraphics[width=3in]{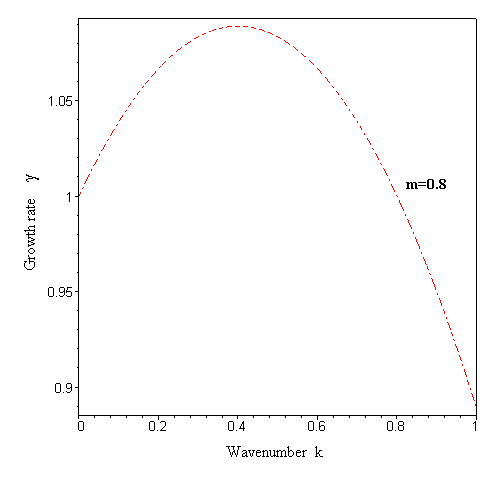}
\end{center}
\begin{center}
Fig.2-4.  The dependence of the growth rate on the wavenumber for
$m=0.1, 0.5, 0.8$.
\end{center}

\end{document}